\documentclass[12pt, draftclsnofoot, onecolumn]{IEEEtran}

\usepackage[utf8]{inputenc}
\usepackage{amsmath}
\usepackage{pdfpages}
\usepackage{amsfonts}
\usepackage{amssymb}
\usepackage{mathtools}
\usepackage{amsthm}
\usepackage{afterpage}
\usepackage{verbatim}
\usepackage{algorithm}
\usepackage{algpseudocode}
\usepackage{stfloats}
\usepackage{cite}
\usepackage{balance}
\usepackage{multirow}
\usepackage{graphicx}
\usepackage{bm}
\usepackage{caption}
\usepackage{subfigure}
\usepackage{xcolor}

\newcommand{\E}{\operatorname{E}}
\newcommand{\I}{\operatorname{I}}

\newcommand{\pr}{\operatorname{P}}

\begin{document}

\title{Optimal Mobility and Communication Strategy \\
to Maximize the Value of Information \\in IoT Networks\\
}
\author{Zijing~Wang,~\IEEEmembership{Graduate~Student~Member,~IEEE,}
        Mihai-Alin~Badiu,~\IEEEmembership{Member,~IEEE,} and~Justin~P.~Coon,~\IEEEmembership{Senior~Member,~IEEE}
\thanks{This research was funded in part by EPSRC grant number EP/T02612X/1 and Clarendon Fund Scholarships at the University of Oxford. For the purpose of Open Access, the author has applied a CC BY public copyright licence to any Author Accepted Manuscript (AAM) version arising from this submission.
}
\thanks{The authors are with the Department of Engineering Science, University of Oxford, Oxford OX1 3PJ, U. K. (e-mail: zijing.wang@eng.ox.ac.uk, mihai.badiu@eng.ox.ac.uk, and justin.coon@eng.ox.ac.uk).}
}

\maketitle
\vspace{-0.5cm}
\begin{abstract}
The Internet of Things (IoT) is an emerging next-generation technology in the fourth industrial revolution. In industrial IoT networks, sensing devices are largely deployed to monitor various types of physical processes. They are required to transmit the collected data in a timely manner to support real-time monitoring, control and automation. The timeliness of information is very important in such systems. Recently, an information-theoretic metric named the ``value of information" (VoI) has been proposed to measure the usefulness of information. In this work, we consider an industrial IoT network with a set of heterogeneous sensing devices and an intelligent mobile entity. The concept of the value of information is applied to study a joint path planning and user scheduling optimisation problem. We aim to maximise the network-level VoI under mobility and communication constraints. We formulate this problem as a Markov decision process (MDP), and an efficient algorithm based on reinforcement learning is proposed to solve this problem. Through numerical results, we show that the proposed method is able to capture the usefulness of data from both time and space dimensions. By exploiting the correlation property of the data source, the proposed method is suitable for applications in resource-limited networks.
\end{abstract}


\section{Introduction}
\IEEEPARstart{B}{y integrating} smart sensing devices, massive machine-type communications (mMTC) and Internet of Things (IoT) technologies, significant changes are taking place in the fourth industrial revolution~\cite{IoTSurvey,IoTsurveyJ}. In the era of industrial 4.0, the main objectives are to achieve real-time monitoring, precise control, decentralised decision-making and reach a high level of automation. An industrial IoT network generally comprises massive sensing devices and mobile intelligent entities (or robots) that sample and analyse data to enable autonomous industrial control. With abilities of wireless communication, self-monitoring and configuration, IoT sensors are installed to continuously monitor different types of complex physical processes, such as temperature, emergency traffic and so on. They are required to transmit the sampled data to the intelligent robot as quickly as possible in order to make sure that the entity has the latest status updates of the monitored process. The entity moves and collects data from a variety of data sources, and it is able to analyse and use the received data for further decision-making without requiring human intervention. Compared with traditional wireless sensor networks (WSNs) in which sensors transmit data via multiple hops to the base station, the mobile robot is able to operate in the best position to adapt to the communication environment by controlling its moving path which further helps to improve the network efficiency.

In such systems, it is of critical importance to optimise the mobility of the robotic entity and communication strategy between sensors and the entity in order to maintain the timeliness of data. The joint path planning and communication optimisation problem has been thoroughly explored with various traditional objectives. The throughput is maximised by optimising the trajectory and resource allocation in orthogonal multiple access (OMA)~\cite{WQQUAVthroughputOFDMA} and non-orthogonal multiple access (NOMA) systems~\cite{UAVthroughputNOMA}, respectively. The completion time is minimised in the unmanned aerial vehicle (UAV)-assisted multicasting sensor networks~\cite{UAVtimeMulticast}. The optimal energy-efficient trajectory is designed in~\cite{UAVenery}. The joint path planning and data collection problem is studied to minimise the estimation error in~\cite{UAVMSE}. These works successfully meet the demand of the traditional wireless sensor networks in which the value of each sampled data packet does not change over time. For real-time IoT applications, data packets are sampled continuously and they are temporally correlated with each other, which means that a newly generated packet should be more important than an old packet. However, these traditional performance measurements are not able to capture the correlation from the time dimension, thus they are not suitable to be employed in time-critical IoT networks.

To measure the freshness of data, the age of information (AoI) has been proposed as a new performance metric in~\cite{2012infocom?,BookConceptTool}. It is defined as the time duration since the generation time of the latest received data packet, and it quantifies the temporal dynamics of data from the destination's perspective. Due to its novelty, the AoI has been widely utilised as an optimisation tool in IoT networks to support real-time control. In single-user networks, the AoI-optimal sampling strategy at the data source is studied with various underlying random processes and various queue models~\cite{QueueSYMin,DistributionISITJournal}. In multi-user networks, the AoI-optimal user scheduling and link activation problems are studied in~\cite{scheduleMinMax} and~\cite{mineWCL}, respectively. Moreover, the AoI-oriented mobility and communication design problem has also received much attention~\cite{UAV2018Info,UAVgrid,UAVicc2019,mineUAV,MengyingSunUAV,UAVflytime,UAVAOIWCSP}. The authors in~\cite{UAV2018Info} firstly explore two path planning problems with the aim of minimising the maximum and average AoI, respectively, in UAV-assisted data collection networks. Dynamic programming (DP) and genetic algorithm (GA) methods are utilised to solve the proposed two problems. The authors in~\cite{UAVgrid} consider the grid-based trajectory to optimise the AoI. This problem is formulated as a Markov decision process (MDP) which is solved by deep reinforcement learning (RL). Energy consumption has been further taken into consideration in the trajectory design, and AoI-energy-aware data collection schemes are proposed in~\cite{mineUAV,MengyingSunUAV}. The authors in~\cite{UAVflytime} jointly optimise the AoI, moving time as well as user association for data acquisition in sensor networks. 

These existing works have successfully captured the time correlation of data samples generated by a single source. However, as the AoI is a linear function of time in which the slope is $1$, the AoI is not able to capture the non-linear performance degradation caused by the physical correlation properties of the data source. For example, the data generated by an emergency traffic monitor changes more frequently than the data sampled by a temperature surveillance device, thus two packets of the same age contain different levels of valuable information. In some cases, old samples of less-changeable data sources may still be useful, but new samples of highly-changeable data sources may hold little useful information. This means that the AoI is not suitable to measure the network-level data freshness in the system with heterogeneous data sources. 

Against the reason mentioned above, there is a rising trend in investigating non-linear data freshness metrics. A common method is to use non-linear AoI functions to measure the timeliness of data~\cite{QueueSYUpdatWait,NonlinearVoUDJ,binaryCache2021,NonlinearEH}. The notion of ``age penalty" is introduced in~\cite{QueueSYUpdatWait}. It is defined as a general non-decreasing AoI function and is interpreted as how dissatisfied the outdated data is to the network. The authors in~\cite{NonlinearVoUDJ} explore how to choose penalty functions based on the autocorrelation of the underlying process. They suggest that the logarithmic penalty function is appropriate for highly-correlated data samples while the exponential function is appropriate for less-correlated samples. The authors in~\cite{binaryCache2021} applied the binary penalty function in cache update networks in which the data sources vary slowly with time. The authors in~\cite{NonlinearEH} derive closed-form expressions of exponential and binary penalty functions with energy harvesting transmitters in various queue models. 

Instead of choosing penalty functions subjectively, information theory-based and signal processing theory-based metrics are also introduced to provide the mathematical basis. The mean squared error (MSE) for remote estimation systems has been extensively investigated for data usefulness~\cite{estimationMarkovSource,estimationOUprocess,estimationWienerProcess,estimationContext-aware}. The relationship between the AoI and MSE is studied in~\cite{estimationMarkovSource}, in which the authors prove that the MSE can be written as a non-decreasing penalty function when the data source is Markov. The MSE-based penalty function is used in two specific Markov random processes and the optimal sampling strategy is obtained~\cite{estimationOUprocess,estimationWienerProcess}. Moreover, the mutual information is used to measure how valuable the information is~\cite{previousGlobecom,SPAWC,previousEntropy,MImultisources,previousCL,NonlinearSurveySY}, in which the authors show that the mutual information can also be regarded as a special age penalty function. The relationship between the AoI and the mutual information is investigated in various Markov processes, e.g., the autoregressive process~\cite{SPAWC} and the Ornstein-Uhlenbeck process~\cite{previousEntropy}. The optimal sampling problem is studied in satellite-integrated Internet of Things networks with multiple sensing devices~\cite{MImultisources}. The relationship between the mutual information-based metric and the MSE-based metric is investigated in~\cite{previousCL}. The data freshness has been greatly treated in Markov models, in which the data samples are assumed to be transmitted successfully once the data source is scheduled for transmission. However, a practical IoT network is very likely to be affected by noise, interference or fading. Therefore, the sample received via a poor wireless channel contains less value than the sample transmitted under good channel conditions. 

Against this background, we proposed a new concept of ``value of information" (VoI) in our previous work~\cite{previousTIT}. The VoI is defined in the context of information theory which can be used to measure the usefulness of information in a noisy environment with heterogeneous data sources. We also derived its closed-form expression in general noisy Gauss-Markov models, and studied its connection to traditional AoI and MSE metrics~\cite{previousCL}. It is shown that the small AoI and small MSE can lead to a large value of information. Therefore, the VoI is a relevant metric which is worth to be optimised for the usefulness and timeliness of information. In this work, we apply this VoI metric in an industrial IoT network and explore the VoI maximisation problem by jointly optimising the mobility and communication strategy. We consider a network with a set of heterogeneous sensing devices and a mobile central entity. Each sensor monitors a specific physical process and the entity wishes to collect the sampled data for further analysis. The goal is to develop a mobility-communication strategy between the sensing devices and the central entity such that the value of information is maximised. Specifically, we aim to find the optimal mobility path of the central entity and the scheduling scheme of each user to maximise the minimum VoI. This VoI optimisation problem is modelled as a Markov decision process, and we propose an intelligent algorithm based on reinforcement learning to solve this problem.

The main contributions of this work are summarised as follows.
\begin{itemize}
    \item An optimisation problem is formulated for valuable data collection in IoT networks with different types of sensing devices. We jointly consider the path planning of the central entity and the communication scheme to maximise the minimum value of information. This problem is formulated as a mixed integer non-linear programming (MINLP) problem with mobility, scheduling, and transmission constraints.
    \item As the proposed problem is intractable, we reformulate it as an MDP and propose a competitive algorithm based on reinforcement learning to solve it efficiently. We obtain the optimal moving direction, time slot allocation and user scheduling scheme so that the network-level value of information at the entity is maximised. 
    \item Numerical results are presented to evaluate the performance of the proposed algorithm. We show that the proposed scheme is able to measure the data value in temporal and spatial domains. Compared with other existing methods, the proposed method achieves better performance when the wireless resource is limited by exploiting the correlation properties of sensing devices.
\end{itemize}

The remainder of this paper is presented as follows. The system model and problem formulation are given in Section II. The problem is reformulated and a reinforcement learning-based algorithm is proposed in Section III. Simulation results and performance analysis are provided in Section IV. Conclusions are drawn in Section V.

\section{System Model and Problem Formulation}

\begin{figure}
\centering
\includegraphics[width=9.5cm]{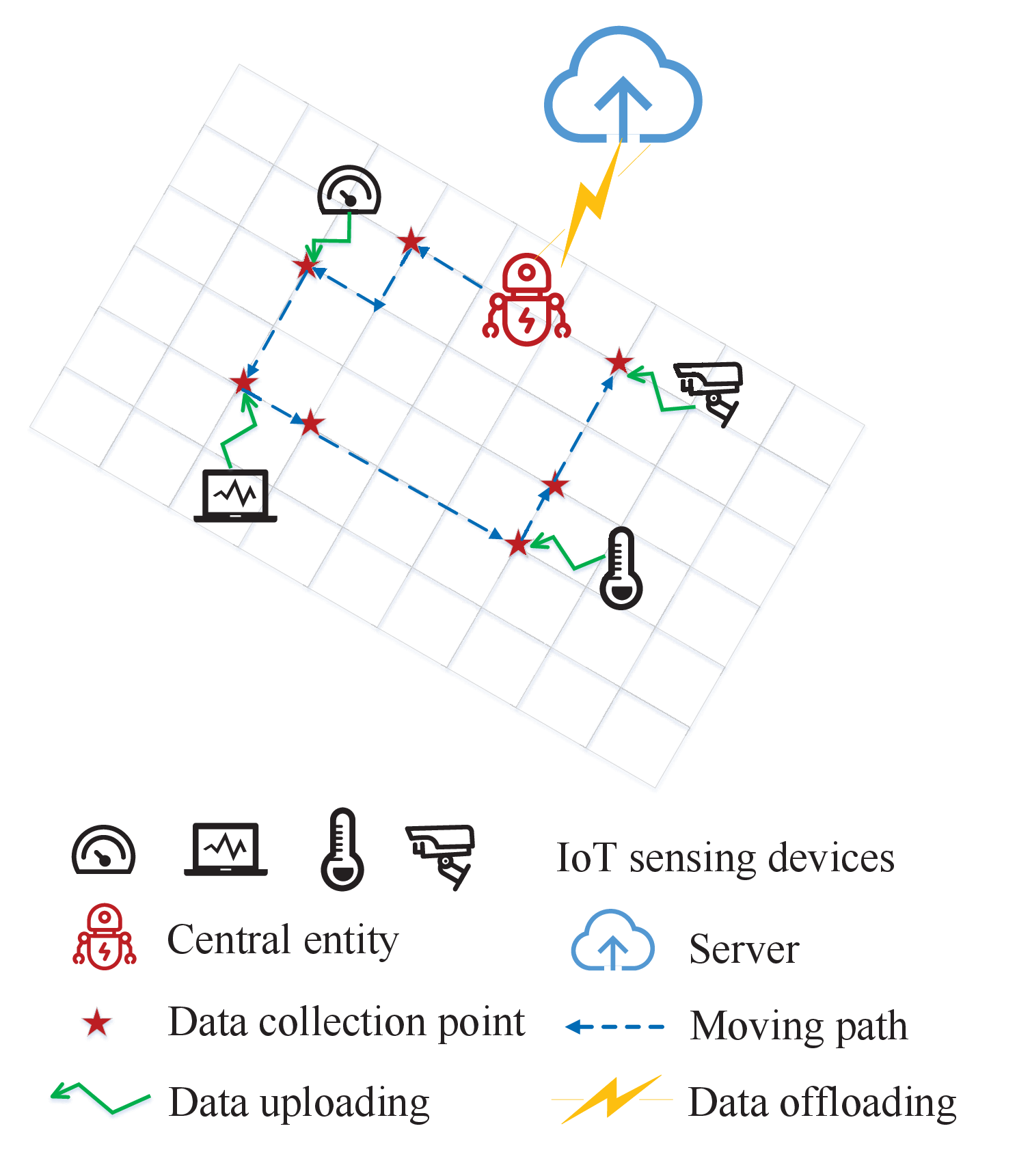}
\caption{System model.}
\label{fig:model}
\end{figure}

As shown in Fig.~\ref{fig:model}, we consider a wireless network with a set of fixed sensing nodes and a mobile central entity. In a practical industrial IoT network, the sensing nodes represent various IoT devices, and the central entity can be used to represent an intelligent robot or a smart vehicle. Various sensing nodes are deployed to monitor the dynamics of heterogeneous physical processes, such as temperature, emergency traffic, light intensity and so on. They are required to upload the status update of the target random process to the central entity in a timely manner and then the entity offloads the data to the server for further analysis to achieve real-time surveillance and precise control. Due to the limited wireless resources (e.g., energy and spectrum), long-distance transmission can lead to large path loss, thus it is difficult to transmit the data packet to the central entity via a long-distance link. Therefore, the central entity needs to move to the best position to fulfil the communication demand by controlling its moving path. To achieve real-time industrial automation, the usefulness of the data at the central entity plays an important role. How to plan the path and schedule the nodes is a key problem in such networks. 

\begin{table}
    \centering
    \renewcommand\arraystretch{1.5}
    \begin{tabular}{ |c|l| } 
        \hline
        Symbol & Definition \\
        \hline
        $N$ & Total number of sensing nodes \\ 
        \hline 
        $T$ & Total number of time slots \\ 
        \hline
        $x_i,y_i$ & Position of the sensing node $i$ \\ 
        \hline
        $x(t),y(t)$ & Position of the central entity at time slot $t$\\
        \hline
        $d_i(t)$ & Distance between the sensing node $i$ and the central \\
        & entity at time slot $t$\\
        \hline
        $q_i(t)$ & Effective distance between the sensing node $i$ and the\\
        & central entity at time slot $t$\\
        \hline
        $z$ & Moving distance of the central entity within a unit time slot \\
        \hline
        $m(t)$ & Moving direction of the central entity at  time slot $t$\\ 
        \hline
        $\eta_i(t)$ & SNR between node $i$ and the central entity at time slot $t$ \\
        \hline
        $P_i$ & Transmission power of node $i$ \\
        \hline
         $\alpha_i(t)$ & A binary variable which indicated whether the node $i$ \\
         &is scheduled to transmit data at time slot $t$ \\
        \hline
        $A_i(t)$ & Age of information of node $i$ at time slot $t$\\
        \hline
        $V_i(t)$ & Value of information of node $i$ at time slot $t$\\
        \hline
        $\rho_i$ & Correlation parameter of node $i$ \\
        \hline
    \end{tabular}
    \caption{Notations}
    \label{tab:my_label}
\end{table}

\subsection{Mobility Model}
 The notations are given in Table I. We consider a network with $N$ fixed sensing devices and one mobile central entity. This network is assumed to be time-slotted and we denote $T$ as the total number of time slots which means that the time is discrete. For simplicity, the duration of each slot is assumed to be $1$. The location of each sensor is denoted by $(x_i,y_i)$ with $1 \le i \le N$, which is assumed to be fixed and known. The location of the mobile central entity within time slot $t$ is denoted by $(x(t),y(t))$. The moving path can be obtained by connecting the location of each separate line segment. The distance between the node $i$ and the central entity at time slot $t$ is denoted as $d_i(t)$ which is given as
\begin{equation}
    \label{eq:distance d_n}
    d_i(t)= \sqrt { {\bigg(x_i-x(t)\bigg)}^2+{\bigg(y_i-y(t)\bigg)}^2}, \quad    (1 \le i \le N, 1 \le t \le T).
\end{equation}

For the central entity's mobility, we assume it moves with a fixed speed and denote $z$ as the moving distance within a unit time slot. $m(t)$ represents the moving direction of the central entity at time slot $t$, and there are five options which can be chosen by the central entity.  $\{\theta_{N},\theta_{S},\theta_{E},\theta_{W}\}$ represents north, south, east and west directions, respectively. If $m(t)=\theta_{0}$, the central entity will retain its current position. Specifically, the location of the central entity is presented by
\begin{equation}
  \label{eq: moving direction}
    \bigg(x(t+1),y(t+1)\bigg)= \left\{ {\begin{array}{*{20}{l}}
    {\bigg(x(t),y(t)+z\bigg),}&{m(t)=\theta_{N}}\\
    {\bigg(x(t),y(t)-z\bigg),}&{m(t)=\theta_{S}}\\
    {\bigg(x(t)+z,y(t)\bigg),}&{m(t)=\theta_{E}}\\
    {\bigg(x(t)-z,y(t)\bigg),}&{m(t)=\theta_{W}}\\
    {\bigg(x(t),y(t)\bigg),}&{m(t)=\theta_{0}}
\end{array}} \right. \quad (1 \le t < T).
\end{equation}

\subsection{Communication Model}
\begin{figure}
\centering
\includegraphics[width=8.5cm]{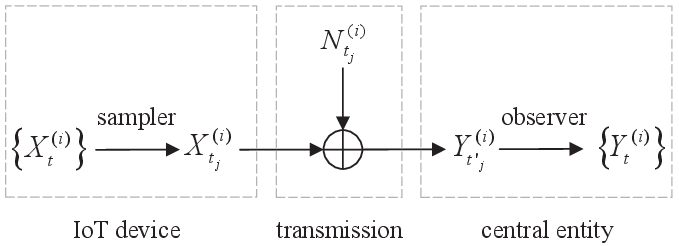}
\caption{Transmission Model.}
\label{fig:flow chart}
\end{figure}

The transmission model is given in Fig.~\ref{fig:flow chart}. For each sensing node, we denote $\{X^{(i)}_t\}$ as the random process under observation at node $i$ with $1 \le i \le N$. Denote $X^{(i)}_{t_j}$ as the data packet which is sampled by the sensor $i$ at time $t_j$. It represents the status of the target physical process $\{X^{(i)}_t\}$ at time $t_j$. We assume the sampled packet is transmitted through a noisy channel and received by the central entity at time $t'_j$. We denote $Y^{(i)}_{t'_j}$ as the corresponding noisy observation which is captured by the entity. As time goes by, the observations are recorded in the observing random process $\{Y^{(i)}_t\}$ at the entity. Due to the existence of transmission delay and transmission noise, the data packet is corrupted. Thus we have $t'_j>t_j$, $Y^{(i)}_{t'_j} \neq  X^{(i)}_{t_j}$ and $\{Y^{(i)}_{t}\} \neq  \{X^{(i)}_{t}\}$. Since we consider the discrete-time system, the time instants here are integers. The data packets are assumed to be transmitted through an additive Gaussian noise channel, and thus we have
\begin{equation}
    Y^{(i)}_{t'_j}=X^{(i)}_{t_j}+N^{(i)}_{t_j}.
\end{equation}
Here, $N^{(i)}_{t_j}$ is the Gaussian noise sample which is independent of the data sample $X^{(i)}_{t_j}$, and we denote $N_0$ as the power spectral density of the Gaussian noise. 

Compared with the urban outdoor network, the quality of transmission in an industrial IoT is less likely to be affected by channel
fading caused by blocks from buildings, and mountains~\cite{LoSIoT}. We assume that sensor devices are able to communicate with the central entity via the line of sight (LoS) link with additive white Gaussian noise, thus the channel condition largely depends on the transmission distance. Based on the free-space path loss model, the signal-to-noise ratio (SNR) between the node $i$ and the central entity at time slot $t$ can be given as
\begin{equation}
\label{eq:snr gamma_n(t)}
    \eta_i(t)=\alpha_i(t)\frac{P_ig_0d_i^{-2}(t)}{BN_0},\quad (1 \le i \le N, 1 \le t \le T)
\end{equation}
where $P_i$ is the transmission power of node $i$, $g_0$ is the channel power gain at the reference distance $d_0=1 \rm{m}$, and $B$ is the available bandwidth. $\alpha_i(t)$ is a binary variable indicating the scheduling policy. If the node $i$ transmits its data successfully at time slot $t$, then $\alpha_i(t)=1$, and vice versa. In this work, we consider the time-division multiple access (TDMA)-based user scheduling scheme and this means that at most one node can be served for a given time slot. Therefore, the time slot allocation constraints can be given as
\begin{equation}
\label{eq:scheduling constraint beta over all sensors}
    \sum\limits_{i = 1}^N {{\alpha_i}(t) }\le 1,\quad (1 \le t \le T).
\end{equation}

\subsection{Value of Information Model}
For any sensing node, its age of information at a given time instant $t$ is defined as the time elapsed since the generation time of the latest observed data packet at the receiver~\cite{2012infocom?,AoISurveyJSAC}, i.e.,
\begin{equation}
\label{eq:AoI def}
    A(t)=t-t_n, \quad t>t'_n.
\end{equation}
Here, $n$ is the total number of observations captured by the entity until time $t$ which is also the index of the most recently received data sample. $t_n$ is the generation time at the source and $t'_n$ is the receiving time at the destination. The AoI measures the temporal correlation of a target process, but it is independent of the physical properties of the underlying random process $\{X^{(i)}_t\}$ and the quality of the observed process $\{Y^{(i)}_t\}$. As illustrated in Fig.~\ref{fig:model}, a practical IoT network is occupied with various sensing devices. Different physical processes have different correlation properties and they update at different speeds over time. Therefore, we are motivated to propose the concept of the value of information for use in our study.

The value of information is a new performance metric which measures the usefulness of data from the mobile central entity's perspective~\cite{previousTIT}. For a given time instant $t$ and a given device $i$, the VoI is defined as the mutual information between the current status of the underlying random process at the transmitter $X^{(i)}_t$ and the latest noisy observation at the receiver $Y^{(i)}_{t'_n}$, i.e.,
\begin{equation}
\label{eq:voi def}
    V_i(t)=\I(X^{(i)}_t;Y^{(i)}_{t'_n}), \quad t>t'_n.
\end{equation}
This VoI concept is interpreted as the reduction in the uncertainty of the current unobserved status given the latest noisy observation. The entropy of $X^{(i)}_t$ reflects the physical property of the underlying random process. Timestamps $t$, $t_n$ and $t'_n$ reflect the data usefulness from the time dimension. The entropy of $Y^{(i)}_{t'_n}$ reflects the level of the transmission noise. Therefore, the VoI is able to capture the timeliness of data, the correlation property of the data source and the channel quality. This means that it is an appropriate metric to evaluate network performance.

In this work, we model the physical process under observation $\{X^{(i)}_t\}$ as a stationary Gauss-Markov random process. It is able to generalise various random processes~\cite{introOU}, such as the discrete autoregressive process and continuous Ornstein–Uhlenbeck process, and it is widely used to model some practical applications in physical sciences. For example, it can represent the length of a spring in the presence of thermal fluctuations in an IoT-assisted monitoring system~\cite{OUspring}. 

The random process $\{X^{(i)}_t\}$ is assumed to be standard Gaussian distributed. To specify such a process, it is sufficient to specify its autocorrelation function~\cite{bookCorrelation}. We denote $\rho_i$ ($0< \rho_i <1$) as the correlation parameter of node $i$, then the autocorrelation between two random variables $X^{(i)}_{t}$ and $X^{(i)}_{\tau}$ is given as
\begin{equation}
    \E\bigg[X^{(i)}_{t}X^{(i)}_{\tau}\bigg]=\rho_i^{|t-\tau|}.
\end{equation}
This correlation parameter represents the correlation of the underlying random process. Specifically, small $\rho$ means less correlation and large $\rho$ means high correlation. 

In this model, the VoI (in bits/s) of the node $i$ at time $t$ in~\eqref{eq:voi def} can be given by~\cite{previousCL,previousTIT}
\begin{equation}
     \label{eq:general voi 1 dimension}
    V_i(t)  = - B\log \bigg(1 - \frac{P_i g_0 {\rho_i ^{A_i(t)}}}{{BN_0q_i^2(t)+P_ig_0 }}\bigg),\quad (1 \le i \le N, 1 \le t \le T).
\end{equation}
Here, $A_i(t)$ represents the AoI of node $i$ at time $t$ which is defined in~\eqref{eq:AoI def}. $q_i(t)$ represents the ``effective distance'' between the node $i$ and the central entity at time slot $t$. In the following part, we will consider two cases, and show how $q_i(t)$ relates to $d_i(t)$ and the AoI evolution.
\begin{itemize}
    \item Case 1: $\eta_i(t) \ge \eta_{th}$. If the SNR is larger than a given threshold $\eta_{th}$, we assume that the transmission is successful. In this case, we have $q_i(t)=d_i(t)$ and $A_i(t)=1$. If the transmission at time $t$ is successful, the effective distance is equivalent to the Euclidean distance. The AoI will be updated to $1$ when a new packet is received by the central entity, and the VoI can be updated accordingly. 
    
    \item Case 2: $\eta_i(t) < \eta_{th}$. If the SNR is smaller than the given threshold, we assume that the transmission is unsuccessful and we need to discard the packet. In the case with network outage, we have $q_i(t)=q_i(t-1)$ and $A_i(t)=A_i(t-1)+1$. This is because the central entity does not receive any effective information regarding the status of the underlying random process, and it can only hold the information at the previous time slot. Therefore, the unsuccessful status transmission will not cause an update but a decrease in the VoI.
\end{itemize}
Thus, the AoI and the effective distance constraints are given as
\begin{equation}
\label{eq: AoI}
     {A_{i}}(t+1) = \left\{ {\begin{array}{*{20}{l}}
{1,}&{\eta_i(t) \ge \eta_{th}}\\
{{A_{i}}(t) + 1,}&{\eta_i(t) < \eta_{th}}
\end{array}} \right.,
 \quad (1 \le i \le N, 1 \le t < T).
\end{equation}
and
\begin{equation}
\label{eq:effective distance}
    {q_i}(t) = \left\{ {\begin{array}{*{20}{l}}
{d_i(t),}&{\eta_i(t) \ge \eta_{th}}\\
{{q_{i}}(t-1) ,}&{\eta_i(t) < \eta_{th}}
\end{array}} \right.,
 \quad (1 \le i \le N, 1 \le t \le T).
\end{equation}

In~\eqref{eq:general voi 1 dimension}, the AoI of each device $A_i(t)$ measures the freshness of data from the time domain. The effective transmission distance $q_i(t)$ measures the timeliness of data from the space domain. The correlation parameter $\rho_i$ reflects the physical properties of the underlying random process. Therefore, the proposed VoI notion is able to capture the usefulness of data in noisy spatial-temporal networks.

\subsection{Problem Formulation}
Our objective is to find the optimal central entity's path planning and the node scheduling scheme to maximise the minimum VoI among all sensor nodes. This max-min optimisation problem is formulated as follows:
\begin{equation*}
\begin{aligned}
\mathop {\max }\limits_{\mathbf{m,\alpha}} \quad & J\\
\text{s.t.}\quad &
{{V_i}(t)} \ge J, \quad (1 \le i \le N, 1 \le t \le T)\\
&\text{Mobility constraints: }~\eqref{eq:distance d_n},~\eqref{eq: moving direction}\\
&\text{Communication constraints: }~\eqref{eq:snr gamma_n(t)},~\eqref{eq:scheduling constraint beta over all sensors}\\
&\text{VoI constraints: }~\eqref{eq:general voi 1 dimension},~\eqref{eq: AoI},~\eqref{eq:effective distance}\\
&\alpha_i(t) \in \{0, 1\},\quad (1 \le i \le N, 1 \le t \le T)\\
&m(t) \in \{\theta_N,\theta_S,\theta_E,\theta_W,\theta_0\},\quad (1 \le t \le T)
\end{aligned}
\end{equation*}

In this problem formulation, ${V_{i}}(t)$, $q_i(t)$, $d_i(t)$, and $\eta_i(t)$ are continuous variables. $A_i(t)$, $m(t)$, $x(t)$ and $y(t)$ are discrete variables. $\alpha_i(t)$ is a binary variable. $T$, $N$, $P_i$, $\eta_{th}$, $B$, $N_0$, $\rho_i$, $g_0$ and $z$ are constants. Our objective is to find the optimal path planning of the central entity ($m(t)$) and the time slot allocation scheme of each sensor node ($\alpha_i(t)$) at each time slot to maximise the minimum VoI over all the nodes in the network.

\section{A Reinforcement Learning-based Algorithm}
This VoI optimisation problem is formulated as a mixed integer non-linear programming problem. Due to the high dynamics and uncertainty of the network, it is not practical to find the optimal solution using traditional methods. Therefore, we reformulate the problem as a Markov decision process first and then propose a reinforcement learning-based method to solve this problem efficiently.

\subsection{Markov Decision Process}
The MDP is widely used to model the decision-making process of an agent. A general MDP includes four parts, i.e., $\{\mathcal{S},\mathcal{A},\mathcal{R},\mathcal{P}\}$, where $\mathcal{S}$ is the state set, $\mathcal{A}$ is the action set, $\mathcal{R}$ is the reward when moving from one state to another state due to an action, and $\mathcal{P}$ is the corresponding state transition probability.

In our work, the central entity is regarded as the agent, and the corresponding MDP is described as follows. 
\begin{itemize}
    \item State: For each time slot $t$, the state includes the central entity's position, the AoI and the effective distance between the sensor and the central entity, which is given as
    \begin{equation}
        s_t=\bigg\{x(t),y(t),\bm{A}(t),\bm{q}(t)\bigg\}
    \end{equation}
    where
    \begin{equation}
    \begin{aligned}
        \bm{A}(t)&={\bigg[A_1(t), A_2(t),\cdots, A_N(t)\bigg]}^{\operatorname{T}},\\
        \bm{q}(t)&={\bigg[q_1(t), q_2(t),\cdots, q_N(t)\bigg]}^{\operatorname{T}}.
        \end{aligned}
    \end{equation}
    Here, $x(t)$ and $y(t)$ jointly represent the position of the central entity, vector $\bm{A}(t)$ represents the AoI of all nodes, and vector $\bm{q}(t)$ represents the effective distance of all nodes. 
    \item Action: For each time slot $t$, the action includes the moving direction and the user scheduling scheme, which is given as
    \begin{equation}
        a_t=\bigg\{m(t),\bm{\alpha}(t)\bigg\}
    \end{equation}
    where 
    \begin{equation}
    \begin{aligned}
        m(t)&\in{\bigg\{\theta_N, \theta_S, \theta_E,\theta_W,\theta_0\bigg\}},\\
        \bm{\alpha}(t)&={\bigg[\alpha_1(t), \alpha_2(t),\cdots, \alpha_N(t)\bigg]}^{\operatorname{T}}.
    \end{aligned}
    \end{equation}
    Here, $m(t)$ represents the moving direction of the agent, and $\bm{\alpha}(t)$ represents the scheduling policy of all nodes at time slot $t$ with $\alpha_i(t) \in \{0,1\}$ for all $1 \le i \le N$.
    \item Reward: For each time slot $t$, the reward relates to the state and action at the previous time slot, which is defined as the minimum VoI among all nodes, i.e.,
    \begin{equation}
    \label{eq:reward}
        r_{t}(s_{t-1},a_{t-1})=\min \{V_1(t), \cdots, V_N(t)\}-f_{penalty}.
    \end{equation}
    Here, $f_{penalty}$ is a non-negative and constant parameter which is used to represent the penalty. Specifically, if no node is visited for a given time slot, we take $f_{penalty}=100$; otherwise, $f_{penalty}=0$. Due to the existence of this penalty parameter, the learning process can be more efficient by stimulating the central agent to collect data as timely as possible and avoiding it stopping at a position with no information being collected. The VoI can be calculated by $\bm{A}(t)$ and $\bm{q}(t)$. Since the total number of time slots $T$ is assumed to be finite, the AoI of each node is finite. Therefore, the VoI of each node in~\eqref{eq:general voi 1 dimension} is also finite.
    \item Transition probability: The state transition probability is defined as the probability of the state in the next time slot $(t+1)$ given the state and action in the time slot $t$, which is denoted as $\pr(s_{t+1}|s_t,a_t)$. 
\end{itemize}
For any given time slot, the MDP is in a state $s_t$ and the central entity needs to choose a feasible action $a_t$ from the available set $\mathcal{A}_{s_t}$. For the following time slot, the MDP moves to the next state $s_{t+1}$ and the central entity receives the related reward $r_{t+1}(s_t,a_t)$. The objective of the MDP is to find a policy that can be used by the central entity to choose how to move and serve sensor nodes for each time slot.

We denote $\pi(a|s)$ as the stationary policy which is a distribution of actions given states, i.e.,
\begin{equation}
    \pi(a|s)=\pr(a_t=a|s_t=s).
\end{equation}
The optimal policy can be obtained by maximising the state-value function. Particularly, the state-value function is defined as the expected long-term discounted reward function which is given as~\cite{MDP}
\begin{equation}
\label{eq:v_pi(s)}
    \mathbb{V}_\pi(s)= \E_{\pi} \bigg[\sum\limits_{\tau = t}^{\infty}{\gamma^{\tau}}{r_{\tau+1}(s_\tau,a_\tau) \mid s_t=s}\bigg].
\end{equation}
$\E_\pi[\cdot]$ represents the expectation taken over different decision-making given different states. Here, $\gamma$ is the discount factor ($0 \le \gamma \le 1$) which measures the level of the importance of future states given the current state. Based on Bellman's expectation equation~\cite{MDP}, the state-value function can be presented by the immediate reward plus the discounted value of the future state, i.e.,
\begin{equation}
\begin{aligned}
    \mathbb{V}_\pi(s) &= \E_{\pi} \bigg[r_{t+1}(s_t,a_t)+\gamma \mathbb{V}_\pi(s_{t+1}) \mid {s_t=s}\bigg]\\
    &=\E_{\pi} [r_{t+1}(s_t,a_t)]+ {\gamma}\sum\limits_{s' \in \mathcal{S}}{\pr(s'|s,a)\mathbb{V}_\pi(s')}.
\end{aligned}
\end{equation}
Here, $s$ and $a$ represent the current state and action. $s'$ represents the next state when action $a$ is taken.

We denote $\pi^*(a|s)$ as the optimal policy which can lead to the optimal state-value function, i.e.,
\begin{equation}
\begin{aligned}
\label{eq:optima v}
     \mathbb{V}_{\pi^*}(s)&=\mathop {\max }\limits_{\pi} \mathbb{V}_{\pi}(s)\\
    &=\mathop {\max }\limits_{a \in \mathcal{A}_s}\bigg[\E_{\pi} [r_{t+1}(s_t,a_t)]+ {\gamma}\sum\limits_{s' \in \mathcal{S}}{\pr(s'|s,a)\mathbb{V}_\pi(s')}\bigg].
\end{aligned}
\end{equation}
We introduce a state-action-value function $\mathbb{Q}_{\pi}(s,a)$ as right hand side of~\eqref{eq:optima v} and it is defined as
\begin{equation}
    \mathbb{Q}_{\pi}(s,a)=\E_{\pi} [r_{t+1}(s_t,a_t)]+ {\gamma}\sum\limits_{s' \in \mathcal{S}}{\pr(s'|s,a)\mathbb{V}_\pi(s')}.
\end{equation}
Thus, we can write
\begin{equation}
\label{eq:pi star}
     \mathbb{V}_{\pi^*}(s)=\mathop {\max }\limits_{a \in \mathcal{A}_s}\mathbb{Q}_{\pi^*}(s,a).
\end{equation}
The optimal state-value function $\mathbb{V}$ can be obtained by the state-action-value function $\mathbb{Q}$ which is given as
\begin{equation}
\label{eq:MDP q recursive}
    \mathbb{Q}_{\pi^*}(s,a)=\E_{\pi} [r_{t+1}(s_t,a_t)]+ {\gamma}\sum\limits_{s' \in \mathcal{S}}{\pr(s'|s,a)\mathop {\max }\limits_{a' \in \mathcal{A}_{s'}}\mathbb{Q}_{\pi^*}(s',a')}.
\end{equation}
An algorithm is proposed in the following section to find the solution $\pi^*$ of our formulated problem.

\subsection{A Q-learning based Algorithm}
Reinforcement learning is a common method to solve the MDP in the context of implicit transition probabilities. Q-leaning is an important and lightweight RL algorithm which can learn the reward of action in a given state with random transition probability based on past experience.

The aim of a Q-learning task is to find the optimal action $a_t$ based on the current state $s_t$. Specifically, the central entity has a Q-table to store the Q-value which relates to the current position and the information value. The central entity will choose an available action based on its Q-value, and the Q-table will be updated after moving to the next state. Through update iterations, the optimal Q-value can be derived eventually in a recursive way.

By implementing the Q-learning algorithm, the optimal Q-value can be achieved by an iterative update based on the following recursion based on~\eqref{eq:MDP q recursive}
\begin{equation}
\label{eq:recursive Q}
     Q_{t+1}(s_{t},a_{t}) \leftarrow (1-\beta)Q_t(s_t,a_t)
    +\beta r_{t+1}(s_t,a_t)+\beta \gamma\mathop {\max }\limits_{a \in {\mathcal{A}_{s_{t+1}}}}Q_t(s_{t+1},a)
\end{equation}
where $\beta$ is the learning rate ($0 <\beta \le 1$) and $\gamma$ is the discounted factor. The updated Q-value relates to three parts. The first term $(1-\beta)Q_t(s_t,a_t)$ represents the current Q-value. The second term $\beta r_{t+1}(s_t,a_t)$ reflects the reward when action $a_t$ is taken in state $s_t$. The third term $\beta \gamma\mathop {\max }\limits_{a \in {{\mathcal{A}}_{s_{t+1}}}}Q_t(s_{t+1},a)$ measures the optimal future reward which can be obtained in state $s_{t+1}$. Based on~\eqref{eq:pi star}, the optimal action is chosen that maximises the Q-value.

\begin{algorithm}
\renewcommand{\algorithmicrequire}{\textbf{Initialisation:}}
\caption{A Q-learning based Algorithm}\label{algo}
\begin{algorithmic}[1]
\Require learning rate $\beta$, discounted factor $\gamma$, Q-table, running rounds $R_{\max}$, total time slots $T_{\max}$, parameters $\epsilon$ and $\epsilon_0$
\For {$i=1:1:R_{\max}$}
\State set initial state $s_1$
\For {$t=1:1:T_{\max}$}
\State $U_{\max}=-1$
\State Generate a random number $x$ ranging from $[0,1]$ 
\If {$x<\epsilon$}
\State {Randomly select an action $a_t \in {\mathcal{A}_{s_t}}$}
\Else
\For {each action $a \in {\mathcal{A}_{s_t}}$}
\If{$Q(s_t,a) > U_{\max}$ }
    \State $U_{\max} = Q(s_t,a)$
    \State $a_t=a$ 
\EndIf
\EndFor
\EndIf
\State {Adopt action $a_t$}
\State {Record the next state $s_{t+1}$ and calculate the reward $r(s_t,a_t)$} based on~\eqref{eq:reward}
\State Update the Q-table based on~\eqref{eq:recursive Q}
\EndFor
\State $\epsilon \leftarrow \epsilon-\epsilon_0$
\EndFor
\end{algorithmic}
\end{algorithm}

The details of the proposed Q-learning-based method are presented in Algorithm 1. First, we need to initialise the following constant parameters: the learning rate $\beta$, the discounted factor $\gamma$, the total number of running rounds $R_{\max}$ and the total number of time slots $T_{\max}$. Q-value in the Q-table is set to an arbitrary but fixed value before running the algorithm. $\epsilon$ and $\epsilon_0$ represent the exploration and exploitation probabilities. The central entity can exploit the learning and estimate the Q-value based on the recursive method to maximise the VoI. However, at the beginning of the algorithm, the central entity does not have the chance to visit all states, so it has no knowledge of whether the estimated value is optimal. Therefore, it is important to explore the next available state-action pair for a given state. In the proposed algorithm, an $\epsilon$-greedy method is utilised to balance exploration and exploitation. The exploitation phase means that the entity chooses the available action that maximises the Q-value. The exploration phase means that the entity chooses a random action from the available action set which is independent of the Q-value. For a given state $s$, the action is selected by the following
\begin{equation}
         a = \left\{ {\begin{array}{*{20}{l}}
{\mathop {\arg \max }\limits_{a \in {\mathcal{A}_{s}}}Q(s,a),}&{\text{with probability } 1-\epsilon }\\
{\text{a random action in } {\mathcal{A}_{s}},}&{\text{with probability } \epsilon.}
\end{array}} \right.\\
\end{equation}
The central entity adopts the selected action, records the next state, calculates the corresponding reward based on~\eqref{eq:reward}, and updates the Q-table based on~\eqref{eq:recursive Q}. For each running round, the probability $\epsilon$ is deceased by $\epsilon_0$ at a fixed rate. As the learning goes on, it will decrease to $0$ and this means that the agent is able to exploit the learning to take optimal action instead of exploration. 
\section{Numerical Results}

\subsection{Simulation Setup}
In this section, numerical results are provided to show the efficiency of the proposed algorithm. We compare the proposed method with the other two solutions to evaluate their performance. They are described as followed.
\begin{itemize}
\item ``VoI-optimal" scheme denotes the proposed method in this paper.
\item ``AoI-optimal" scheme in which the objective function is to minimise the maximum age of information among all nodes. Similarly to the proposed VoI-optimal scheme, we also employ reinforcement learning to solve the problem. The difference is that the minimum VoI is replaced by the negative maximum AoI in the design of the reward function.
\item ``Shortest path" scheme in which the objective is to minimise the total length of the mobility path. This scheme can be used as the benchmark to compare the performance between different methods.
\end{itemize}

In the simulation, the fixed sensor nodes are deployed randomly in the $5 \times 5$ area. The starting position of the mobile central entity is also selected randomly. The simulation parameters are given as followed. The channel power gain $g_0$ at the reference distance $d_0=1\rm{m}$ is set to $-50 \rm{dB}$. The available bandwidth $B$ is set to $2 \times 10^6 \rm{Hz}$. The noise power spectral density $N_0$ is set to $-110 \rm{dBm/Hz}$. The moving distance per time slot is set to $z=1$. We assume that all nodes have the same transmission power. In this case, the SNR threshold $\eta_{th}$ only depends on the transmission distance. We use $d_{th}$ to represent the distance threshold of the successful transmission and set $d_{th}$ as $1$. The learning rate $\beta$ is set to $0.75$, and the discount factor $\gamma$ is set to $0.9$. The total running round $R_{\max}$ is set to $10^6$, the initial exploration probability $\epsilon$ is set to $1$ and exploitation probability $\epsilon_0$ is set to $10^{-6}$. 

\begin{figure}
    \centering
    \subfigure[The VoI-optimal scenario.]
    {
        \includegraphics[width=8.5cm]{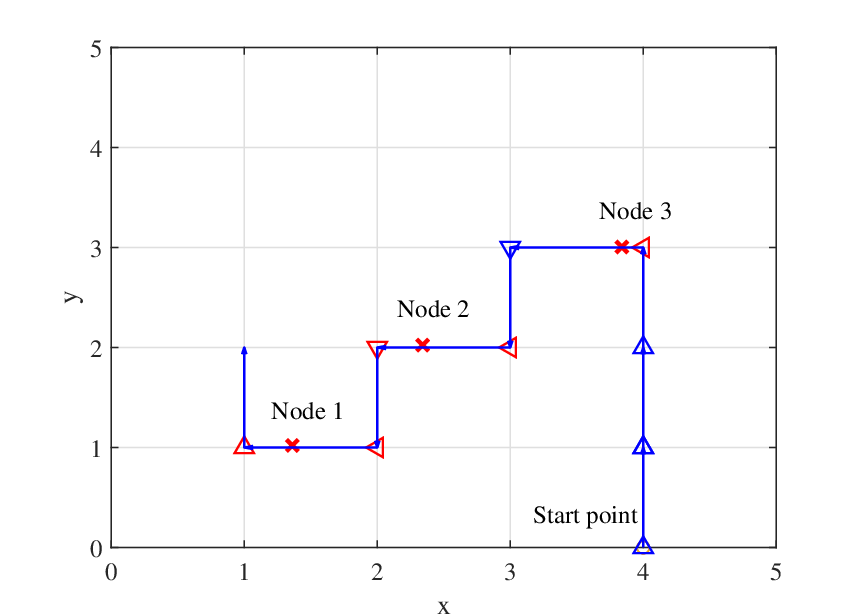}
        \label{fig:CaseVoI}
    }
    \subfigure[The AoI-optimal scenario.]
    {
        \includegraphics[width=8.5cm]{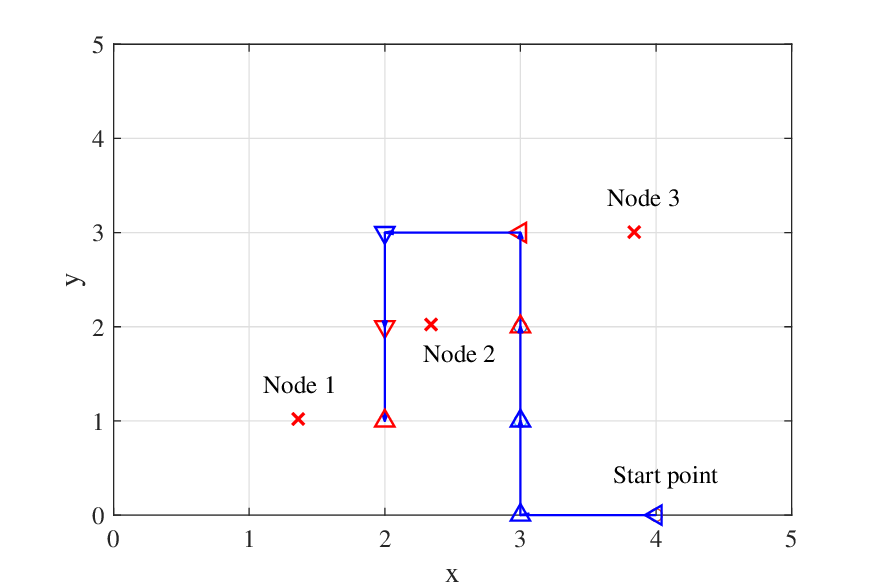}
        \label{fig:CaseAoI}
    }
    \subfigure[The shortest path scenario.]
    {
        \includegraphics[width=8.5cm]{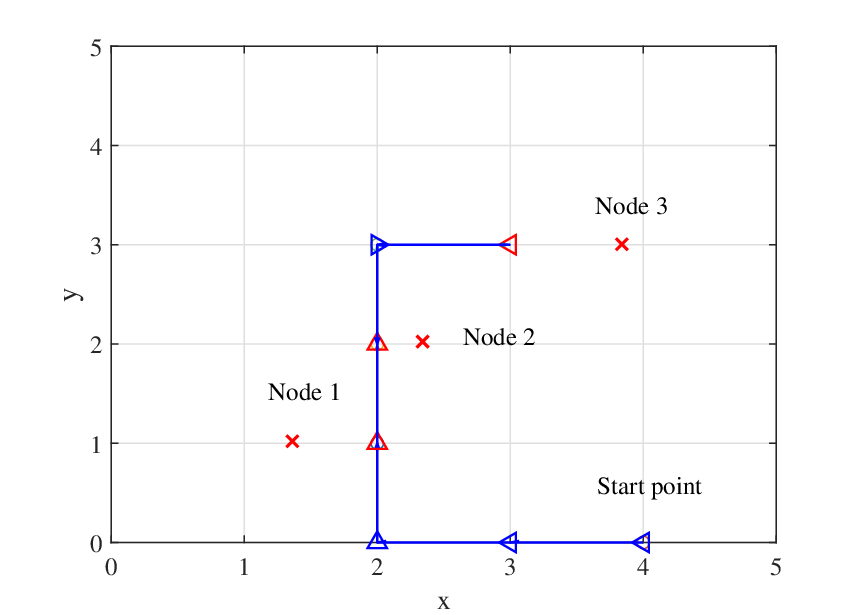}
        \label{fig:CaseShortest}
    }
    \caption{A case study. The number of nodes $N=3$, the correlation parameter of each node $\rho=[0.7,0.6,0.5]$, initial age of each node $A=[2,4,4]$, the transmission power of each node $P=1\rm{W}$, and the total number of time slots $T_{\max} = 10$.}
    \label{fig:casestudy}
\end{figure}

Fig.~\ref{fig:casestudy} gives a case study to illustrate how the proposed algorithm and the benchmark method work in the same network. Fig.~\ref{fig:vs Time} further evaluates and compares the network performance in this given network scenario. Figs.~\ref{fig:vs power} to~\ref{fig:NodeNum} illustrate the average network performance in terms of the VoI, in which $40$ network instances are generated randomly for each network setting.

\subsection{Performance Analysis}

Fig.~\ref{fig:casestudy} gives three examples to show the mobility of the central entity and the communication scheme between the sensing nodes and the central entity by three different solutions. The sensing devices are fixed and marked by the ``red cross". The position of the central entity at each time slot is marked by the ``triangle" and the trajectory is marked by the ``blue line". The direction of the triangle represents the moving direction and the colour represents the communication indicator. ``Red triangle" means that a node $i$ is scheduled to communicate with the central entity at this position (i.e., $\alpha_i(t)=1$); otherwise, the triangle is blue. For example, in Fig.~\ref{fig:CaseVoI}, the central entity moves to $(4,3)$ at time slot $4$. In this position, node $3$ transmits its data to the central entity and the entity will move to $(3,3)$ at the next slot. 

\begin{figure}
\centering
\includegraphics[width=9.5cm]{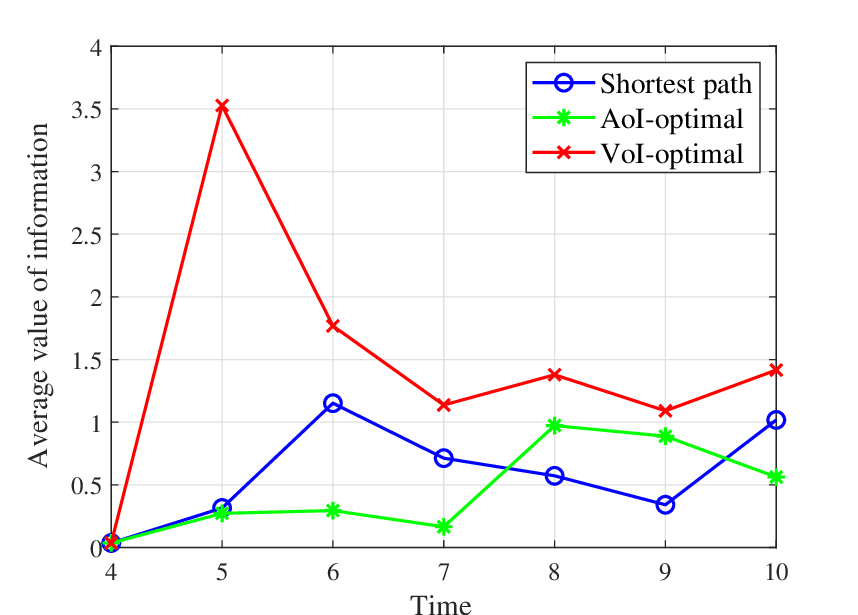}
\caption{The average value of information evolution among all nodes in the network given in Fig.~\ref{fig:casestudy}.}
\label{fig:vs Time}
\end{figure}

The solution of the proposed algorithm is visualised in Fig.~\ref{fig:CaseVoI}, and the AoI-optimal and shortest path schemes are shown in Figs.~\ref{fig:CaseAoI} and~\ref{fig:CaseShortest}, respectively. Fig.~\ref{fig:vs Time} shows how the average VoI evolves with time from the central entity's perspective, illustrating that the proposed algorithm achieves better performance. This is because the VoI metric is able to capture the timeliness of data and the correlation properties of data. In this network setting, node $1$ is highly correlated, node $2$ is intermediately correlated and node $3$ is less-correlated. Therefore, node $3$ varies more quickly with time thus it should be firstly served with priority. However, the shortest path solution only considers the moving distance, and the AoI-optimal solution does not take the correlation into consideration. Moreover, in Fig.~\ref{fig:CaseVoI}, the data collection position of node $3$ is $(4,3)$ while this position is $(3,3)$ in the other two schemes. This means that the proposed VoI-optimal solution is able to capture the performance degradation caused by the different transmission distances. Therefore, this case study illustrates that the VoI metric reflects the correlation properties, and the timeliness of data in both time and space domains.

\begin{figure}
\centering
\includegraphics[width=9.5cm]{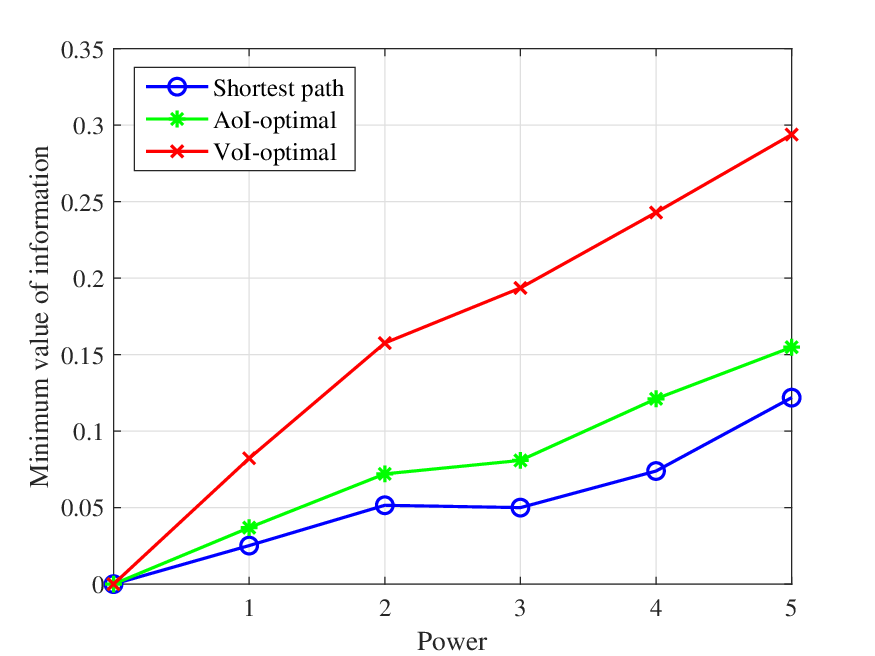}
\caption{The time average minimum value of information among all nodes versus the transmission power. The number of nodes $N=4$, the correlation parameter is generated uniformly from $[0,1]$, and the total number of time slots $T_{\max}=30$.}
\label{fig:vs power}
\end{figure}

Fig.~\ref{fig:vs power} presents how the minimum value of information varies with the transmission power. This figure shows that the VoI increases with the transmission power, and this means that the VoI is able to capture the SNR in the transmission environment and verifies that the packet received via a good channel is more valuable than the packet received via a poor channel. However, the performance of the traditional AoI-optimal or shortest-path schemes is regardless of the transmission distance, thus they are unable to capture the usefulness of data from the spatial dimension.

\begin{figure}
\centering
\includegraphics[width=9.5cm]{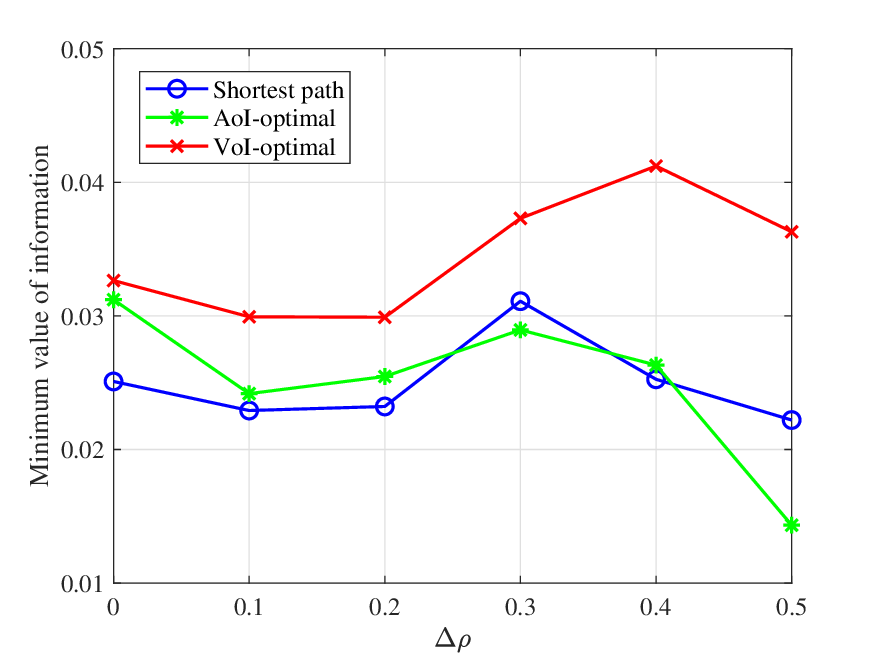}
\caption{Time average minimum value of information among all nodes versus different correlation parameters. The number of nodes $N=4$, the transmission power of each node $P=1\rm{W}$, and the total number of time slots $T_{\max}=30$.}
\label{fig:hetersrc}
\end{figure}

Fig.~\ref{fig:hetersrc} explores the performance of the VoI in the context of a network with heterogeneous data sources. The correlation parameter of each node is generated uniformly from $(0.5-\Delta \rho,0.5+\Delta \rho)$. The horizontal axis represents how heterogeneous the data sources are. Large $\Delta \rho$ represents high heterogeneity; small $\Delta \rho$ represents less heterogeneity. The vertical axis is the time average minimum VoI among all nodes. When $\Delta \rho=0$ (the data sources are homogeneous), the VoI can be written as a monotonic function of the AoI, thus the VoI-optimal solution achieves similar performance as its AoI counterpart. However, their performance gap increases with $\Delta \rho$. This is because the proposed method is more likely to serve the less-correlated node to balance the VoI. But the other two methods do not consider the correlation property and they are not able to help improve the VoI. This figure further shows that our proposed method is more appropriate to be applied in the network with different types of data sources.

\begin{figure}
\centering
\includegraphics[width=9.5cm]{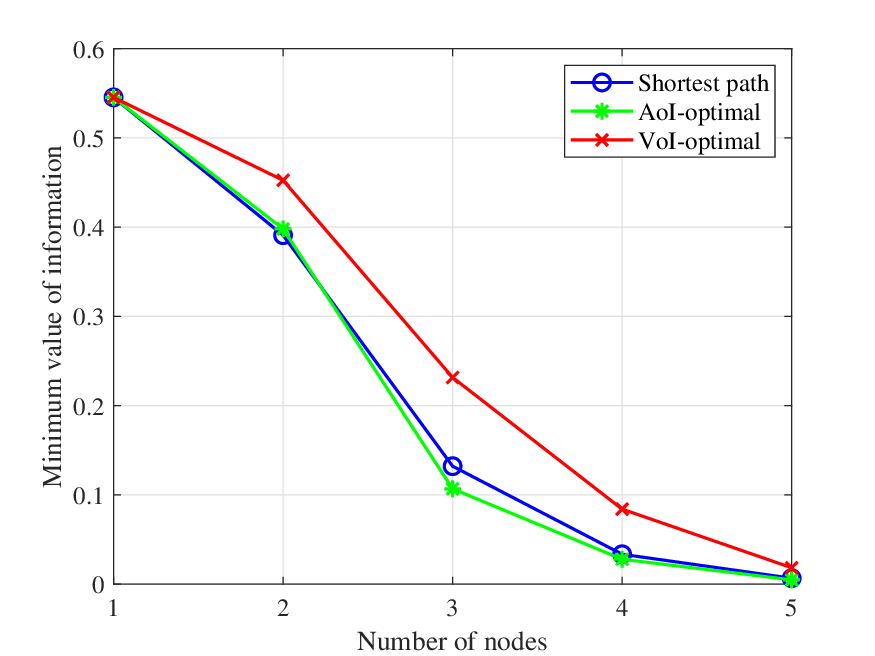}
\caption{Time average minimum value of information among all nodes versus the number of nodes. The correlation parameter is generated uniformly from $[0,1]$, the transmission power of each node $P=1\rm{W}$, and the total number of time slots $T_{\max}=30$.}
\label{fig:NodeNum}
\end{figure}

Fig.~\ref{fig:NodeNum} shows the performance of the minimum value of information versus the number of nodes. When the number of nodes is small, all three methods have similar performance. In this case, the wireless resource and the time slots are enough to serve the node as timely as possible. When increasing the number of nodes, the advantage of the proposed method becomes more and more significant. This is because the VoI is able to explore the correlation properties of the data source, thus less-correlated sources can be served with higher priority. Compared to the shortest path or the AoI-optimal scheme, the proposed method is complex in the algorithm design, but it is more suitable to be applied in networks with limited wireless resources by considering the correlation of the underlying data source.

\section{Conclusions}
In order to achieve real-time industrial control, timely data acquisition is of great importance in IoT networks. In this paper, we apply the concept of the value of information in industrial IoT networks to explore the trajectory design and data collection problem. Our objective is to maximise the minimum value of information by optimising the path planning and user scheduling strategy. This problem is formulated as an MDP, and we propose a reinforcement learning-based algorithm to solve this problem more efficiently. Numerical results are presented to compare the proposed solution with other existing methods. We have shown that the proposed method is able to capture the usefulness of data from both time and space dimensions. We have also shown that, by exploiting the correlation property of the data source, the proposed method is competitive in the network with limited wireless resources. 

\bibliographystyle{IEEEtran}
\bibliography{IEEEabrv,papers}

\end{document}